\documentstyle[11pt]{article}

\title{\bf On the inclusive gluon jet production off the nucleus in the
perturbative QCD}
\author{M.A.Braun\\
Dep. of High Energy physics,
 University of S.Petersburg,\\
198904 S.Petersburg, Russia}
\def\beq{\begin{equation}}
\def\eeq{\end{equation}}
\def\noi{\noindent}
\begin{document}
\maketitle
\medskip
\noi{\bf Abstract.}
In the perturbative QCD approach
single and double inclusive cross-sections for gluon production
off the nucleus  are studied from  the relevant reggeized gluon
diagrams. Various terms  corresponding to emission of gluons from
the triple Pomeron vertex are found. Among them the term derived
by Kovchegov and Tuchin emerges as a result of transition from
the diffractive to effective high-energy vertex. However it
does not exhaust all the vertex contributions to the inclusive
cross-section. In the double inclusive cross-section a
contribution violating the naive AGK rules is found, in which one gluon
is emitted from the vertex and the other from one of the two Pomerons
below the vertex. But then this contribution is subdominant at high energies
and taking it into account seems to be questionable.

\vspace{0.5cm}

\section{Introduction}
After the equation for the BFKL pomeron in the nucleus (BK equation)
had been written,
analyzed analyticaly and solved numerically ~\cite{kovch,bra1}
 the corresponding
inclusive rate of gluon jet production was studied in ~\cite{bra2}
 on the basis
of the AGK rules. The obtained cross-sections follow from the cut
upper pomeron in the fan diagrams for the whole amplitude
(Fig. 1 $a$), the contribution from the cut lower pomerons
(Fig. 1 $b$) cancelled
by the AGK rules. Obviously the resulting cross-section is
linear in the sum of the pomeron fan diagrams $\Phi$.
Some time ago   Yu.Kovchegov
and K.Tuchin derived the same inclusive rate of gluon jet production
in the colour dipole formalism ~\cite{KT}.
Their result is different from ~\cite{bra2} and
corresponds to the substitution in the expression obtained in ~\cite{bra2}
\beq  2\Phi\to 2\Phi-\Phi^2 \eeq
with a negative quadratic term in $\Phi$.
The new term corresponds to emission of gluons from the 3P vertex
itself, which is not prohibited by the AGK rules (Fig.1 $c$).
This gives us a motivation to
reconsider the derivation of the inclusive jet production rate
in the fan diagram formalism, which  is basically much more
transparent in searching for  real intermediate states observed as
gluon jets, as compared to the colour dipole formalism.
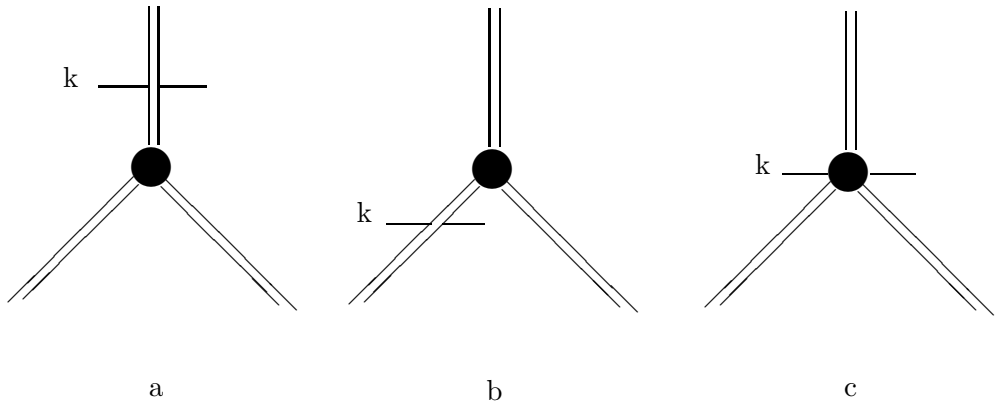
\begin{figure}
\unitlength=1mm
\special{em:linewidth 0.4pt}
\linethickness{0.4pt}
\begin{picture}(147.33,131.67)
\put(35.33,110.33){\circle*{5.20}}
\put(35.00,131.67){\line(0,-1){18.33}}
\put(36.33,131.67){\line(0,-1){18.33}}
\put(33.00,109.00){\line(-1,-1){16.67}}
\put(33.67,108.00){\line(-1,-1){15.33}}
\put(37.33,108.67){\line(1,-1){17.33}}
\put(36.67,107.67){\line(1,-1){15.67}}
\put(80.67,110.00){\circle*{5.20}}
\put(80.33,131.33){\line(0,-1){18.33}}
\put(81.67,131.33){\line(0,-1){18.33}}
\put(78.33,108.67){\line(-1,-1){16.67}}
\put(79.00,107.67){\line(-1,-1){15.33}}
\put(82.67,108.33){\line(1,-1){17.33}}
\put(82.00,107.33){\line(1,-1){15.67}}
\put(128.00,109.67){\circle*{5.20}}
\put(127.67,131.00){\line(0,-1){18.33}}
\put(129.00,131.00){\line(0,-1){18.33}}
\put(125.67,108.33){\line(-1,-1){16.67}}
\put(126.33,107.33){\line(-1,-1){15.33}}
\put(130.00,108.00){\line(1,-1){17.33}}
\put(129.33,107.00){\line(1,-1){15.67}}
\put(28.33,121.00){\line(1,0){6.67}}
\put(36.33,121.00){\line(1,0){6.33}}
\put(66.67,102.67){\line(1,0){6.00}}
\put(74.00,102.67){\line(1,0){5.67}}
\put(119.33,109.33){\line(1,0){6.00}}
\put(131.00,109.33){\line(1,0){6.00}}
\put(24.67,122.33){\makebox(0,0)[cc]{k}}
\put(63.67,104.33){\makebox(0,0)[cc]{k}}
\put(116.67,110.67){\makebox(0,0)[cc]{k}}
\put(36.00,80.67){\makebox(0,0)[cc]{a}}
\put(81.00,80.67){\makebox(0,0)[cc]{b}}
\put(128.33,80.67){\makebox(0,0)[cc]{c}}
\end{picture}
\vspace*{-8cm}
\caption{Pomeron diagrams for the single inclusive cross-section
on two centers}
\label{fig1}
\end{figure}

Our approach is based on  the direct  inspection of reggeized gluon
diagrams, which are assumed to satisfy
the standard AGK rules. They can be divided in two contributions:
from the double Pomeron exchange and from the triple Pomeron interaction
with a certain ("diffractive") 3P vertex $Z$. The study of
the latter contribution allows to localize real
gluons inside the vertex $Z$, which can be observed, and thus find the
inclusive cross-section corresponding to the emission from the vertex $Z$.
Next we pass from the vertex $Z$ to a different vertex $V$, which
incorporates all the contributions at high energies and describes the
splitting of the Pomeron in the non-linear BFKL equation.
This transition
generates a new term which also has  a structure of emission from the vertex.
This term exactly corresponds to the one introduced
by Kovchegov and Tuchin (the KT term).

As a result, our final inclusive cross-section is found to be much richer
than introduced in both our previous paper and by Kovchegov
and Tuchin. Apart from the terms derived in these papers
 it contains several new ones,
also quadratic in $\Phi$ and rather
complicated in structure. The numerical influence of these new terms
will be studied in  following publications.

In the final part of the paper we study the double inclusive cross-section
on the same lines. Here again we find various contributions
corresponding to emisssion from the vertex. Among them there is a
contribution which violates the standard AGK rules. However this violation
is much weaker than claimed in ~\cite{JK}, where also contributions
corresponding to emission from both the upper and lower Pomerons were
found. Also terms which violate the AGK rules are subdominant at high energies,
so that their association with the BK equation is questionable.

\section{Emission from the Pomerons}
In this section, also mainly introductory, we reproduce formulas
for the single and double inclusive cross-sections
which correspond to the emission of gluons from the
Pomerons. This will serve to fix our normalization and compare
with additional contributions  coming from emission  from the 3P vertex.

The scattering amplitude on a nucleus at a given
rapidity $Y$ and impact parameter $b$
is presented as
\beq
{\cal A}(Y,b)=2is\int d^2r\rho(r)\Phi(Y,b,r).
\eeq
Here $\rho(r)$ is the colour dipole density in the projectile and
$\Phi$ is the sum of all Pomeron fan diagrams with the vertex $(1/2)V$,
where $V$ is the effective high-energy 3P vertex introduced in
~\cite{MP,BW,BV}.
Symbolically
\beq
\Phi(r)=P(r)-\frac{1}{2}G(r)VP^2+...,
\eeq
where $P(r)$ is the Pomeron in configuration space, $G(r|r')$ is the
corresponding BFKL  Green function; $G$ and $V$ are assumed to be
operators acting on the Pomeron coordinates and rapidities.
In the following we always
suppress the fixed argument $b$ and often the rapidities when their values
are clear.
Emission from the Pomeron corresponds to "opening" the BFKL chain,
which is described by inserting the emission operator (see ~\cite{BT} and
also Appendix 1.)
\beq
V_k(r)=\frac{4\alpha_sN_c}{k^2} \stackrel{\leftarrow}\Delta e^{ikr}
\stackrel{\rightarrow}\Delta,
\eeq
that is  substituting
\beq
G_Y(r_1|r_2)\to \int d^2r G_{Y-y}(r_1|r)V_k(r)G_y(r|r_2),
\eeq
where we indicated the rapidities as subindeces.

In this way we find the single inclusive cross-section corresponding
to the emission from the uppermost Pomeron in $\Phi$ (at fixed $b$):
\beq
J(y,k)\equiv\frac{(2\pi)^3d\sigma}{dy d^2k d^2b}
=2\int d^2r_1d^2r\rho(r_1)G_{Y-y}(r_1|r)V_k(r)\Phi_y(r).
\eeq
The AGK rules tell us that emission from lower Pomerons in $\Phi$ does
not give any contribution.

The double inclusive cross-section has two contributions from Pomerons.
The first is the double emission from the uppermost Pomeron  (Fig. 2 $a$)
\beq
J_1(y_1,k;y_2,l))=2\int d^2r_1d^2rd^2r'
\rho(r_1)G_{Y-y_1}(r_1|r')V_k(r')G_{y_1-y_2}(r'|r)V_l(r)\Phi_{y_2}(r).
\eeq
The second contribution comes as emissions from the
two Pomerons immediately below the vertex (Fig. 2 $b$)
\[
J_1(y_1,k;y_2,l))=
-\int_{y_1}^Y dy\int d^2rd^2r' \prod_{i=1}^4d^2r_i\rho(r_1)
G_{Y-y}(r_1|r_2)
\]\beq V(r_2|r_3,r_4)
G_{y-y_1}(r_3|r)V_k(r)\Phi_{y_1}(r)G_{y-y_2}(r_4|r')V_l(r')
\Phi_{y_2}(r'),
\eeq
where $V(r_2|r_3,r_4)$ is the 3P vertex in the coordinate space
for the forward direction.

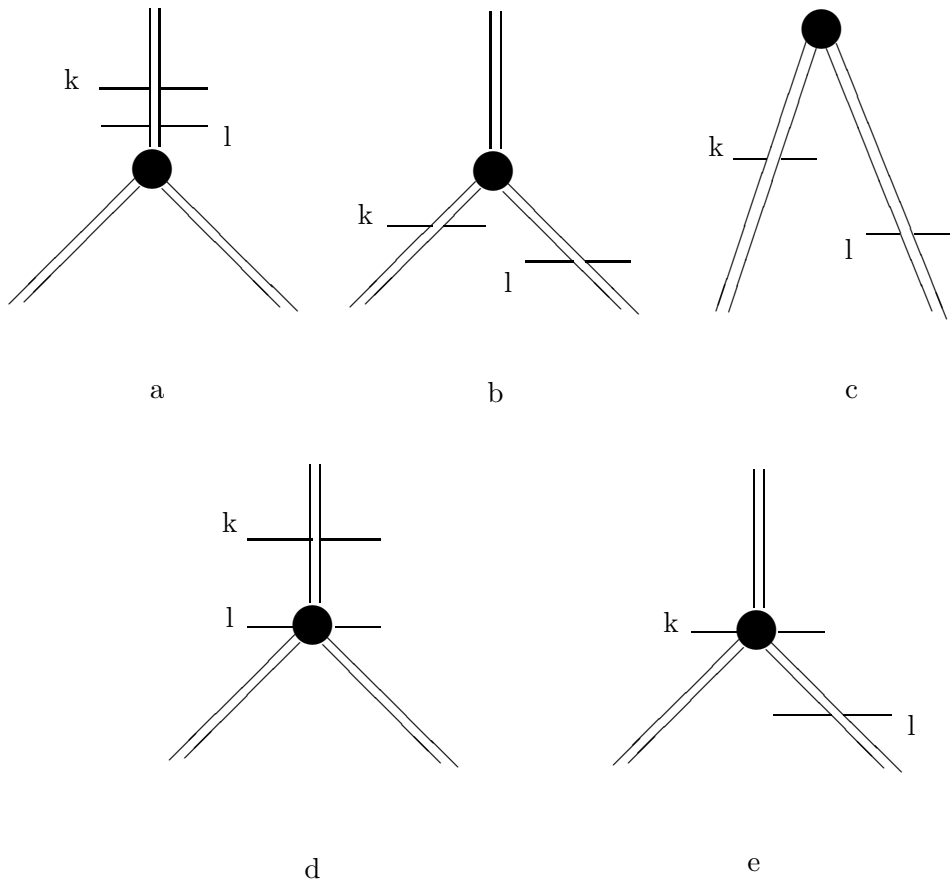
\begin{figure}
\unitlength=1mm
\special{em:linewidth 0.4pt}
\linethickness{0.4pt}
\begin{picture}(141.33,131.67)
\put(35.33,110.33){\circle*{5.20}}
\put(35.00,131.67){\line(0,-1){18.33}}
\put(36.33,131.67){\line(0,-1){18.33}}
\put(33.00,109.00){\line(-1,-1){16.67}}
\put(33.67,108.00){\line(-1,-1){15.33}}
\put(37.33,108.67){\line(1,-1){17.33}}
\put(36.67,107.67){\line(1,-1){15.67}}
\put(80.67,110.00){\circle*{5.20}}
\put(80.33,131.33){\line(0,-1){18.33}}
\put(81.67,131.33){\line(0,-1){18.33}}
\put(78.33,108.67){\line(-1,-1){16.67}}
\put(79.00,107.67){\line(-1,-1){15.33}}
\put(82.67,108.33){\line(1,-1){17.33}}
\put(82.00,107.33){\line(1,-1){15.67}}
\put(115.67,49.00){\circle*{5.20}}
\put(115.33,70.33){\line(0,-1){18.33}}
\put(116.67,70.33){\line(0,-1){18.33}}
\put(113.33,47.67){\line(-1,-1){16.67}}
\put(114.00,46.67){\line(-1,-1){15.33}}
\put(117.67,47.33){\line(1,-1){17.33}}
\put(117.00,46.33){\line(1,-1){15.67}}
\put(28.33,121.00){\line(1,0){6.67}}
\put(36.33,121.00){\line(1,0){6.33}}
\put(66.67,102.67){\line(1,0){6.00}}
\put(74.00,102.67){\line(1,0){5.67}}
\put(107.00,48.67){\line(1,0){6.00}}
\put(118.67,48.67){\line(1,0){6.00}}
\put(24.67,122.33){\makebox(0,0)[cc]{k}}
\put(63.67,104.33){\makebox(0,0)[cc]{k}}
\put(104.33,50.00){\makebox(0,0)[cc]{k}}
\put(36.00,80.67){\makebox(0,0)[cc]{a}}
\put(81.00,80.67){\makebox(0,0)[cc]{b}}
\put(128.33,80.67){\makebox(0,0)[cc]{c}}
\put(28.67,116.00){\line(1,0){6.33}}
\put(36.33,116.00){\line(1,0){6.33}}
\put(85.00,98.00){\line(1,0){6.33}}
\put(93.00,98.00){\line(1,0){6.00}}
\put(56.67,49.67){\circle*{5.20}}
\put(56.33,71.00){\line(0,-1){18.33}}
\put(57.67,71.00){\line(0,-1){18.33}}
\put(54.33,48.33){\line(-1,-1){16.67}}
\put(55.00,47.33){\line(-1,-1){15.33}}
\put(58.67,48.00){\line(1,-1){17.33}}
\put(58.00,47.00){\line(1,-1){15.67}}
\put(48.00,49.33){\line(1,0){6.00}}
\put(59.67,49.33){\line(1,0){6.00}}
\put(45.67,63.33){\makebox(0,0)[cc]{k}}
\put(48.00,61.00){\line(1,0){8.67}}
\put(57.67,61.00){\line(1,0){8.00}}
\put(118.00,37.67){\line(1,0){7.67}}
\put(127.33,37.67){\line(1,0){6.33}}
\put(124.00,129.00){\circle*{0.00}}
\put(124.33,129.00){\circle*{5.20}}
\put(122.33,127.33){\line(-1,-3){12.00}}
\put(123.67,126.33){\line(-1,-3){11.67}}
\put(126.33,127.00){\line(2,-5){14.67}}
\put(125.00,126.33){\line(2,-5){14.00}}
\put(112.67,111.67){\line(1,0){4.33}}
\put(119.00,111.67){\line(1,0){4.67}}
\put(130.33,101.67){\line(1,0){4.33}}
\put(136.67,101.67){\line(1,0){4.67}}
\put(45.33,114.67){\makebox(0,0)[cc]{l}}
\put(82.67,95.33){\makebox(0,0)[cc]{l}}
\put(110.33,113.33){\makebox(0,0)[cc]{k}}
\put(128.00,99.67){\makebox(0,0)[cc]{l}}
\put(136.33,36.33){\makebox(0,0)[cc]{l}}
\put(56.67,17.33){\makebox(0,0)[cc]{d}}
\put(115.33,17.67){\makebox(0,0)[cc]{e}}
\put(45.67,50.67){\makebox(0,0)[cc]{l}}
\end{picture}
\vspace*{-1cm}
\caption{Pomeron diagrams for the double inclusive
cross-section on two centers}
\label {fig2}
\end{figure}

\section{Single inclusive cross-section:
emission from  vertex $Z$}

Contributions presented in the previous sections ("naive")
were introduced and studied in ~\cite {bra2}.
However the analysis of the amplitude with 4 reggeized gluons shows that
apart from  real gluons in the Pomeron chains there appear new
real gluons in the process of changing the number of reggeized gluons or
interactions between different Pomerons. So one may expect additional
contribution to the inclusive cross-sections coming from observation
of real gluons inside the 3P vertex.
To this end it is sufficient to study the scattering on two centers
(nucleons), which corresponds to the second term in the series (3).
Then we can investigate emission
from the vertex as it follows from inspection of
the amplitude for 4 reggeized gluons obtained
in the high-colour limit in ~\cite{bra3}.
Our main idea is that the results for
the amplitude should be consistent with the AGK rules,
which tell that the
relation between the diffractively cut, double cut and single cut
amplitude is $1:2:-4$. The validity of these rules is actually based
only on the fact that the 3P vertex is real and does not change for
various cuttings, which is true in our case.
Knowing this and inspecting particular contributions
we then can establish whether and how  the 3P vertex is cut and from
this find the corresponding contribution to the inclusive cross-section.
Note that the amplitude $D_4$ for 4 reggeized gluons introduced in
~\cite{BW} and studied in ~\cite{bra3}
and below refers to the triple discontinuity
in all energetic variables. The amplitude corresponding to the scattering
on two centers is therefore obtained as $-(1/2)D_4$. This factor $-(1/2)$
appears as a coefficient before the vertex in the summation of
Pomeron fans in (3) and
has to be remembered when translating our formulas from ~\cite{bra3}
into the ones for the inclusive cross-sections.

\subsection{Transitions $2\to 4$ gluons}
We suppress the Pomerons connecting  the amplitude $D_4$ to the
two scattering centers ("attached from below').
For the transition from 2\ to 4 gluons  we have a single contribution to
the amplitude, which we then write as
\beq
D_{2\to4}^{(0)}=-g^4N_cW(1,23,4|1',4')\otimes D(1').
\eeq
We use notations from ~\cite{bra3}. Momenta are denoted by the
number of the gluon
which carry them: $1\equiv k_1$ etc. Notation 12 means the sum of momenta
1 and 2 . Function $W$ is the momentum part of the Bartels kernel
$K_{2\to 3}$:
\beq
W(k_1,k_2,k_3|q_1,q_3)=\frac{(k_1+k_2+k_3)^2}{q_1^2q_3^2}
+\frac{k_2^2}{(k_1-q_1)^2(k_3-q_3)^2}-\frac{(k_1+k_2)^2}{q_1^2(k_3-q_3)^2}
-\frac{(k_2+k_3)^2}{q_3^2(k_1-q_1)^2}.
\eeq
It conserves the momentum so that $k_1+k_2+k_3=q_1+q_3$.
It is assumed that pairs of final gluons 12 and 34 are
colourless and are to be coupled to two final Pomerons. $D(1)$ is the
initial pomeron (amputated). Symbol $\otimes$ in this section
means integration over $1'$
with the weight $(2\pi)^{-3}$.
Diagrammatically contribution (9) is shown in Fig. 3. One has to understand
that in the whole amplitude the gluon pairs 12 and 34  are to be
represented by the standard BFKL ladders.
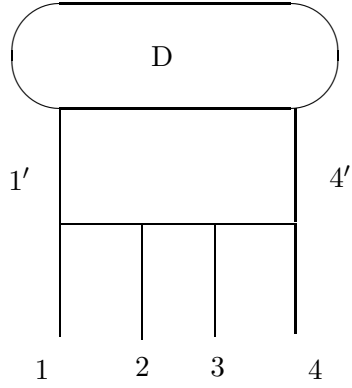
\begin{figure}
\unitlength=1mm
\special{em:linewidth 0.4pt}
\linethickness{0.4pt}
\begin{picture}(77.67,104.00)
\put(55.67,97.00){\oval(43.33,14.00)[]}
\put(40.33,90.00){\line(0,-1){15.33}}
\put(71.67,90.00){\line(0,-1){15.00}}
\put(40.33,74.67){\line(1,0){31.33}}
\put(40.33,74.67){\line(0,-1){15.00}}
\put(71.67,74.67){\line(0,-1){14.67}}
\put(51.33,74.67){\line(0,-1){15.33}}
\put(61.00,74.67){\line(0,-1){15.00}}
\put(38.00,55.33){\makebox(0,0)[cc]{1}}
\put(74.33,55.33){\makebox(0,0)[cc]{4}}
\put(51.33,55.67){\makebox(0,0)[cc]{2}}
\put(61.33,55.67){\makebox(0,0)[cc]{3}}
\put(35.00,80.67){\makebox(0,0)[cc]{$1'$}}
\put(77.67,81.00){\makebox(0,0)[cc]{$4'$}}
\put(54.00,97.00){\makebox(0,0)[cc]{D}}
\end{picture}
\vspace{-5cm}
\caption{Transitions from 2 to 4 gluons}
\label{Fig5}
\end{figure}

Now we pass to the determination of the inclusive cross-section
corresponding to a real gluon inside the 3P vertex.

According to the AGK rules the diffractive contribution is 1/2 of (9).
From Fig. 3 we observe that in the diffractive contribution the central
real gluon is cut in $W$ and the two side ones are uncut. To find the
inclusive cross-section corresponding to observation of the cut central
gluon we have to fix its momentum. Therefore the contribution to
the inclusive cross
section to observe a  gluon of momentum $k$ coming from the vertex will be
\beq
I^{dif}_{2\to 4}=-g^4N_cW(1,23,4|12k,34-k) D(12k).
\eeq
Of course there will be no integration, so that $\otimes$ pases into a
simple product. The factor $(2\pi)^{-3}$ is assumed to be included in $I$.
We have also taken into account factor 2 in (2).

The double cut contribution is just (9) according to the AGK rules.
Since both lower Pomerons are now cut, we conclude from Fig. 3 that both
side gluons in $W$ are cut and the central gluon is uncut. To find the
corresponding inclusive from the vertex we have to fix the momentum of one
of the side gluons. The two possibilities give the same contribution so
that finally
\beq
I^{double}_{2\to 4}=-4g^4N_cW(1,23,4|1k,4-k) D(1k).
\eeq

The single cut contributions give (9) multiplied by $(-2)$. In them only one
of the side gluons in $W$ is cut, the other and the central left uncut.
So to find the inclusive cross-section we have just to fix the momentum
of the cut gluon. We find
\beq
I^{single}_{2\to 4}=4g^4N_cW(1,23,4|1k,4-k) D(1k).
\eeq

    As a result, the double and single cut contributions cancel and
we are left with the diffractive contribution (11), which gives the total
inclusive cross-section from the vertex from transitions $2\to 4$

\subsection{Transitions $3\to 4$ gluons}
According to ~\cite{bra3} the contribution from transitions $3\to 4$ gluons
to the amplitude is given by
\[
D_{3\to4}^{(0)}=g^3\sqrt{2N_c}\Big\{
W(1,2,3|1',3')\otimes D_3^{(134)}(1',3',4)
-W(1,2,4|1',4')\otimes D_3^{(134)}(1',3,4')\]\beq
+W(2,3,4|2',4')\otimes D_3^{(124)}(1,2',4')
-W(1,3,4|1',4')\otimes D_3^{(124)}(1',2,4')\Big\},
\eeq
where
\beq
D_3^{(123)}(1,2,3)=g\sqrt{\frac{N_c}{8}}\Big(D(2)-D(1)-D(3)\Big).
\eeq

Let us consider
the first term in the sum (14) as an example.
It is illustrated in Fig. 4. Again one has
to imagine that the pairs of legs 12 and 34 actually are the beginning of
two BFKL ladders corresponding to lower Pomerons.
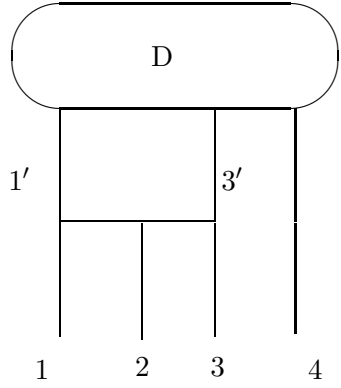
\begin{figure}
\unitlength=1mm
\special{em:linewidth 0.4pt}
\linethickness{0.4pt}
\begin{picture}(77.33,104.00)
\put(55.67,97.00){\oval(43.33,14.00)[]}
\put(40.33,90.00){\line(0,-1){15.33}}
\put(71.67,90.00){\line(0,-1){15.00}}
\put(40.33,74.67){\line(0,-1){15.00}}
\put(71.67,74.67){\line(0,-1){14.67}}
\put(51.33,74.67){\line(0,-1){15.33}}
\put(61.00,74.67){\line(0,-1){15.00}}
\put(38.00,55.33){\makebox(0,0)[cc]{1}}
\put(74.33,55.33){\makebox(0,0)[cc]{4}}
\put(51.33,55.67){\makebox(0,0)[cc]{2}}
\put(61.33,55.67){\makebox(0,0)[cc]{3}}
\put(35.00,80.67){\makebox(0,0)[cc]{$1'$}}
\put(54.00,97.00){\makebox(0,0)[cc]{D}}
\put(61.00,89.67){\line(0,-1){14.67}}
\put(40.33,75.00){\line(1,0){20.67}}
\put(63.33,80.67){\makebox(0,0)[cc]{$3'$}}
\end{picture}
\vspace{-5cm}
\caption{Transitions from 3 to 4 gluons}
\label{Fig6}
\end{figure}

We again start with the diffractive cut. Obviously it corresponds to
the cut right gluon in $W$, while the left one stays uncut. The
contribution to the inclusive cross-section from the vertex will be
\beq
I_{3\to4}^{dif,1}=g^3\sqrt{2N_c}
W(1,2,3|12k,3-k)\otimes D_3^{(134)}(12k,3-k,4).
\eeq
In the double cut amplitude it is the left gluon in $W$ which will be cut,
while the right one will be uncut. Taking into account the factor 2
from the AGK rules we find the contribution from the 1st term in (12):
\beq
I_{3\to4}^{double,1}=2g^3\sqrt{2N_c}
W(1,2,3|1k,23-k)\otimes D_3^{(134)}(1k,23-k,4).
\eeq
Finally in the once cut amplitude the contribution will come only from
the half of all terms, namely from those in which the lower Pomeron
12 is cut. If the Pomeron 34 is cut there is no real gluon in the vertex.
So the total factor is $(-2)$ and the observed gluon from the vertex
is the left one, while the right one is unobserved. The contribution
to the inclusive cross-section from the vertex will be
\beq
I_{3\to4}^{single,1}=-2g^3\sqrt{2N_c}
W(1,2,3|1k,23-k)\otimes D_3^{(134)}(1k,23-k,4).
\eeq

As for transitions $2\to 4$ the contributions from the double cut and single
cut amplitudes cancel and we are left with only the diffractive
contribution (16). The same result holds for the rest of the terms in (14).

\subsection{Transitions $4\to 4$ gluons}
The contribution to the amplitude from transitions $4\to 4$ gluons is
given by \cite{bra3}:
\beq
D_{4\to4}^{(0)}=g^2\Big(U_{23}+U_{14}-U_{13}-U_{24}\Big)\otimes
\Big(D_4^{(1234)}-D_4^{(2134)}\Big),
\eeq
where
\beq
D_4^{(1234)}(1,2,3,4)=\frac{1}{4}g^2N_c\Big(D(1)+D(4)-D(14)\Big)
\eeq
and
\beq
D_4^{(2134)}(1,2,3,4)=\frac{1}{4}g^2N_c\Big(D(2)+D(3)-D(12)-D(13)\Big).
\eeq
Here $U_{23}=U(2,3|2',3')$ is the BFKL interaction
acting between the gluons 2 and 3:
\beq
U(k_1,k_2|q_1,q_2)=\frac{k_1^2q_2^2+k_2^2q_1^2}{q_1^2q_2^2(k_1-q_1)^2}
-\frac{(k_1+k_2)^2}{q_1^2q_2^2},
\eeq
with $k_1+k_2=q_1+q_2$.

As an example we consider the term with $U_{23}$
illustrated in Fig. 5.
\begin{figure}
\unitlength=1mm
\special{em:linewidth 0.4pt}
\linethickness{0.4pt}
\begin{picture}(77.33,104.00)
\put(55.67,97.00){\oval(43.33,14.00)[]}
\put(40.33,90.00){\line(0,-1){15.33}}
\put(71.67,90.00){\line(0,-1){15.00}}
\put(40.33,74.67){\line(0,-1){15.00}}
\put(71.67,74.67){\line(0,-1){14.67}}
\put(51.33,74.67){\line(0,-1){15.33}}
\put(61.00,74.67){\line(0,-1){15.00}}
\put(38.00,55.33){\makebox(0,0)[cc]{1}}
\put(74.33,55.33){\makebox(0,0)[cc]{4}}
\put(51.33,55.67){\makebox(0,0)[cc]{2}}
\put(61.33,55.67){\makebox(0,0)[cc]{3}}
\put(54.00,97.00){\makebox(0,0)[cc]{D}}
\put(61.00,89.67){\line(0,-1){14.67}}
\put(63.33,80.67){\makebox(0,0)[cc]{$3'$}}
\put(51.33,90.00){\line(0,-1){15.00}}
\put(51.33,75.00){\line(1,0){9.67}}
\put(48.00,80.67){\makebox(0,0)[cc]{$2'$}}
\end{picture}
\vspace{-5cm}
\caption{Transitions from 4 to 4 gluons}
\label{Fig7}
\end{figure}
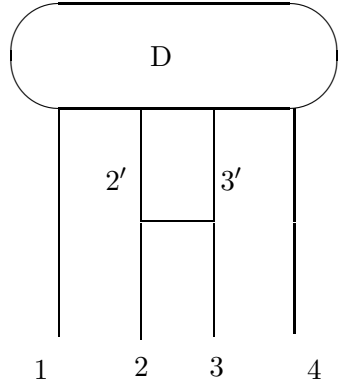
Obviously
both single and double cuts through the lower Pomerons do not pass
through the interaction $U_{23}$, which acts between the two
Pomerons. So the only contribution will come from the diffractive cut.
Fixing the observed gluon momentum we find the contribution to the
inclusive cross-section from the vertex as
\beq
I_{4\to 4}^{(23)}=I_{4\to4}^{dif,(23)}=g^2U(2,3|2k,3-k)
\Big(D_4^{(1234)}(1,2k,3-k,4)-D_4^{(2134)}(1,2k,3-k,4)\Big).\eeq
The total contribution is the sum of this term with the ones coming from
interactions $U_{14}$, $U_{13}$ and $U_{24}$ (with thir respective signs)
which are treated in a similar manner.

\subsection{Total contribution}
We have found that all the contribution $I^{(Z)}$ to the inclusive
cross-section
from the vertex $Z$ only comes from the diffractive cut. Integration over the
observed gluon momentum obviously gives the modulus of the total amplitude
\beq
\int\frac{d^2k}{(2\pi)^3}I^{(Z)}(k)= Z\otimes D.
\eeq
To this expression one should attach two lower Pomerons,
which has been  always assumed implicitly. Then (24) means that the
integrated contribution of the emission from the vertex $Z$
is equal to the total
high-mass diffractive cross-section.

Summing
(11), 4 terms of the type  (16) and 4 terms of the type (23), using
the symmetry in the gluons 12 and 34 and attaching two
lower forward Pomeron fans $\Phi(1)$ and $\Phi(3)$  with $12=34=0$
we find
 \[
I^{(Z)}(k)=g^2N_c
\int\frac{d^2k_1}{(2\pi)^2}\frac{d^2k_3}{(2\pi)^2}
\Phi(1)\Phi(3)\]\[
\Big\{W(1,2,3|12k,3-k)\Big(D(3-k)-D(4)\Big)+
W(2,3,4|2k,34-k)\Big(D(2k)-D(1)\Big)\]\beq-
W(1,23,4|12k,34-k)D(12k)+U(2,3|2k,3-k)\Big(D(13-k)-D(14)\Big)\Big\}.
\eeq
Here we have put $P\to (1/g^2)\Phi$ for the lower Pomerons and
$D\to g^2D$ for the upper one in accordance with their definitions in
~\cite{bra1}.
More explicit expressions for the inclusive cross-section $I^{(Z)}(k)$
can be found in the Appendix 2.

\section{Transformation to the vertex $V$}
The total inclusive cross-section for gluon production on two centers
is a sum of (25)
and a contribution from the cut upper pomeron in the triple Pomeron
diagram with vertex $Z$:
\beq
I^{tot}(k)=I^{(Z)}(k)+J^{(Z)}(k),
\eeq
where we denoted $J^{(Z)}$ the standard contribution (8)
obtained by cutting the upper Pomeron for two scattering centers
with vertex $Z$:
\[
J^{(Z)}(k)=-\frac{g^2N_c}{2\pi}\int\frac{d^2k_1}{(2\pi)^2}
\frac{d^2k_3}{(2\pi)^2}\frac{d^2k'_1}{(2\pi)^2}
 \frac{d^2k_5}{(2\pi)^2}
 \]\beq P(1)P(3)
 Z(1,-1,3,-3|1',-1')\tilde{G}(1'|5)U(5|5k)D(5k).
\eeq
Here $\tilde{G}(1'|5)$ is the forward BFKL Green function, amputated from
the left (that is without the left factor ${k'_1}^{-4}$), satisfying
the equation
\beq
(j-1-2\omega(1)-U)\tilde{G}(1|1')=(2\pi)^2\delta^2(1-1'),
\eeq
where $\omega$ is the gluon trajectory.
We denote
\beq
B(1,2,3,4|5)=G_4(1,2,3,4|1',2',3',4')\otimes Z(1',2',3',4'|5',-5')\otimes
\tilde{G}(5'|5).
\eeq
where $\otimes$ means
integration over all repeated momenta with weight $(2\pi)^{-2}$,
1234=0, and
$G_4(1,2,3,4|1',2',3',4')=G(1,2|1',2')G(3,4|3',4')$.
In fact also $12=1'2'=34=3'4'=0$, however this will be irrelevant
for the time being. Obviously  $B$ satisfies the equation
\beq
S_4B=Z\otimes \tilde{G} +g^2N_c(U_{12}+U_{34})B.
\eeq
This is actually the same equation as in our derivation of the vertex $V$
in ~\cite{BV}, with the difference that $D$ is substituted by $\tilde{G}$.
The latter is also a solution of the BFKL equation although with a
different inhomogeneous term. So we can repeat all our derivation of the
vertex $V$ in \cite{BV} substituting in $D_{40}$
\beq
D_{20}(q)\to \tilde{G}^{(0)}(q|k_5)=(2\pi)^2\delta^2(q-k_5)
\eeq
and separating from $B$ a 'reggeized term'
\beq
B^{R}=D_{40} \Big(D_{20}(q)\to \tilde{G}(q|5)\Big).
\eeq

As a result, we shall get
\[
B=B^R
\]\beq+G_4(1,2,3,4|1',2',3',4')\otimes \Big\{Z(1',2',3',4'|5',-5')\otimes
G_2(5',5)-D_{40}\Big(D_{20}(q)\to \tilde{G}^{(0)}(q|5)\Big)\Big\}.
\eeq
Put into (27) these three terms will give three contributions to the
inclusive cross-section, which have transparent interpretation.
The reggeized term $B^{R}$ attached to the cut interaction $U(5|5k)$
and the upper amputated pomeron $D(5k)$ will give the contribution
corresponding to 'opening' the reggeized  term for the amplitude
$D_4^R$. This term is subdominant at high rapidities and not taken
into account in the non-linear BFKL equation.
The second term will generate the standard contribution from the triple
P interaction with vertex $V$ following from the AGK rules and coming
from the upper pomeron. The third term in (33) is independent of rapidity and
put into (27) will give a contribution $I_1(k)$ which has a meaning
of emission from the vertex.

Explicitly we have
\beq
D_{40}\Big(D_{20}(q)\to \tilde{G}^{(0)}(q|5)\Big)=\frac{1}{2}g^2(2\pi)^2
\Big\{\sum_{i=1}^4\delta^2(i-5)-\sum_{i=2}^4\delta^2(1i-5)\Big\}.
\eeq
From the first term in the first sum we shall have a contribution
\beq
I_1^{(1)}(k)=\frac{1}{2}g^2(2\pi)^2\frac{g^2N_c}{2\pi}
\int\frac{d^2k_1}{(2\pi)^2}\frac{d^2k_3}{(2\pi)^2}
 \frac{d^2k_5}{(2\pi)^2}P(1)P(3)
 U(5|5k)D(5k)\delta^2(1-5).
\eeq
Integration over $k_3$ puts the gluons in the Pomeron $P(3)$ at the same
point in the configuration space and makes the total contribution vanish.
The same argument is true for all terms in the first sum.

The first term in the second sum gives
\beq
I_1^{(2)}(k)=-\frac{1}{2}g^2(2\pi)^2\frac{g^2N_c}{2\pi}
\int\frac{d^2k_1}{(2\pi)^2}\frac{d^2k_3}{(2\pi)^2}
 \frac{d^2k_5}{(2\pi)^2}P(1)P(3)
 U(5|5k)D(5k)\delta^2(12-5).
\eeq
However $12=0$, so that now integrations over $k_1$ and $k_3$  will put
gluons in both the lower pomerons at the same point in configuration space.
So this term also vanishes.

We are left with the two last terms in the second sum. They give equal
contributions and their sum is
\beq
I_1^{(3)}(k)=-g^2(2\pi)^2\frac{g^2N_c}{2\pi}
\int\frac{d^2k_1}{(2\pi)^2}\frac{d^2k_3}{(2\pi)^2}
 \frac{d^2k_5}{(2\pi)^2}P(1)P(3)
 U(5|5k)D(5k)\delta^2(13-5).
\eeq
Using the explicit expression
\beq
U(5,6)=2\frac{k_5^2}{k_6^2(k_5-k_6)^2}
\eeq
and passing to the non-amputated upper pomeron
\beq
D(6)=k_6^4P(6)
\eeq
we rewrite the contribution (37) in the final form
\beq
I_1^{(3)}(k)=-g^2\frac{g^2N_c}{\pi}\frac{1}{k^2}
\int\frac{d^2k_1}{(2\pi)^2}\frac{d^2k_3}{(2\pi)^2}
P(1)P(3)(k_1+k_3)^2(k_1+k_3+k)^2P(13k).
\eeq
Passing to the configuration space one can see that this contribution
is exactly the additional term derived by Kovchegov and Tuchin in the
dipole formalism in ~\cite{KT}. Putting $P\to g^2P$
for the upper Pomeron and $P\to (1/g^2)\Phi$ for the lower ones we find
\beq
I_1^{(3)}(k)\equiv I^{(KT)}(k)=
-\frac{4\alpha_sN_c}{k^2}\int d^2r e^{ikr}\Delta P(r)
\Delta \Phi^2(r).
\eeq
Thus the Kovchegov-Tuchin (KT) term is just the difference between
the standard AGK contributions obtained by cutting the upper Pomeron
in the triple P diagram with verteces $Z$ and $V$.
The interpretation of this term in terms of reggeized gluon diagrams
is illustrated in Fig. 6.

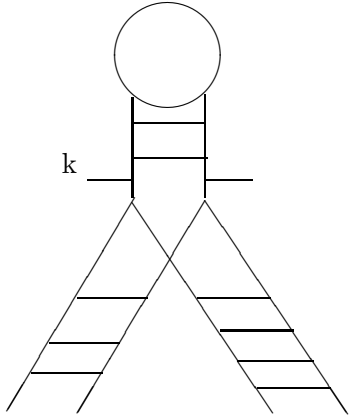
\begin{figure}
\unitlength=1mm
\special{em:linewidth 0.4pt}
\linethickness{0.4pt}
\begin{picture}(98.33,126.00)
\put(74.33,119.00){\circle{14.00}}
\put(69.67,113.33){\line(0,-1){13.33}}
\put(79.33,113.67){\line(0,-1){13.67}}
\put(70.00,100.00){\line(-3,-5){17.00}}
\put(79.33,99.67){\line(-3,-5){17.00}}
\put(69.67,99.33){\line(2,-3){18.67}}
\put(79.33,99.67){\line(2,-3){19.00}}
\put(58.67,80.67){\line(1,0){9.33}}
\put(56.33,76.67){\line(1,0){9.33}}
\put(78.33,86.67){\line(1,0){9.67}}
\put(81.33,82.33){\line(1,0){9.67}}
\put(86.33,74.67){\line(1,0){9.67}}
\put(62.33,86.67){\line(1,0){9.33}}
\put(83.67,78.33){\line(1,0){10.00}}
\put(63.67,102.33){\line(1,0){6.00}}
\put(79.33,102.33){\line(1,0){6.33}}
\put(69.67,105.33){\line(1,0){10.00}}
\put(69.67,110.00){\line(1,0){9.67}}
\put(61.33,104.67){\makebox(0,0)[cc]{k}}
\end{picture}
\vspace{-7cm}
\caption{The Kovchegov-Tuchin term for the inclusive cross-section}
\label {fig6}
\end{figure}

However from our derivation it follows that the total contribution from
the vertex $V$ has to additionally include the part $I^{(Z)}$ which
appears in the direct calculation with vertex $Z$:
\beq
I^{(V)}=I^{(KT)}+I^{(Z)}.
\eeq

\section{Double inclusive cross-section}
Now we pass to the double inclusive cross-section $I(k,l)$ for the production
of two gluon jets with rapidities and momenta $y_1,k$ and $y_2,l$ with
$y_1>>1$,  $y_2>>1$ and $Y>>y_1>>y_2$ where $Y$ is the overall rapidity.
In the following we shall often suppress
rapidities $y_1$ and $y_2$ associating them with the given
observed gluon momenta.
By the AGK rules and without emission from the vertex the double inclusive
cross-section is described by the first three diagrams shown in Fig. 2.
Allowed by the AGK rules emission from the vertex adds to them a new
diagram Fig. 2 $d$. Note that in contrast to the
single inclusive cross-section
different contributions  to the double inclusive cross-section
behave differently at large rapidities.
Two emisssions from the upper Pomeron (Fig. 2 $a$) and one from the
upper Pomeron and the other from the vertex (Fig. 2 $d$) generate cross-
sections which grow as $\exp \Delta(Y+y_2)$,
whereas emissions from the double Pomeron exchange (Fig. 2 $c$)
or the two lower Pomerons  (Fig. 2 $b$)
give contributions which grow as
$\exp 2\Delta Y$ and clearly dominate. So
taking into account the former contributions
(or any with a similar high-rapidity behaviour) in
the summation of Pomeron fans is not justified, as this
summation  is standardly performed in the high-rapidity limit.
This problem is not academic, since  as we shall see, analysis of the
reggeized gluon diagrams will produce new contributions and among them
a contribution  corrresponding to the diagram Fig. 2 $e$
prohibited by the AGK rules
and with the emission from the vertex different from Fig. 2 $d$.
However all of them are subdominant relative to the emission from the double
Pomeron exchange or the two lower
Pomerons (Fig. 2 $a,c$) and strictly speaking have to be neglected.

\subsection{Emission from the vertex}
As with the single inclusive cross-section we begin with the study
of emission from the vertex itself. Since the two observed gluons are
assumed to have widely different rapidities, only one of them may be
emitted from the vertex. Obviously there may be two possibilities.
Either it is the slowest gluon $l$ which is emitted from the vertex,
the quicker one $k$ emitted from the upper Pomeron, or the quicker
gluon $k$ is emitted from the vertex and the slow $l$ from one of
the lower Pomerons.  We start from the first possibility.

If the gluon $l$ is emitted from the vertex and the gluon $k$ is emitted
from the upper Pomeron (Fig. 2 $d$) then the corresponding cross-section
can be found in a trivial manner. In fact in the diagrams for the single
inclusive cross-section the upper Pomeron is always cut. So to find the
cross-section corresponding to Fig. 2 $d$ all we have to do is to
take the single inclusive emission from the vertex  $I^{(Z)}(l)$,
Eq. (25), and "open" the upper Pomeron to describe  emission of the
gluon $k$ from it, that is substitute the upper Pomeron $D(q)$
according to (5), which in the momentun space means
\beq
D(q)\to G(q,q_1)\otimes V_k(q_1,q_2)\frac{D(q_2)}{q_2^4},
\eeq
where
\beq
V_k(q_1,q_2)=\frac{4\alpha_cN_c}{K^2}q_1^2q_2^2(2\pi)^2\delta^2(q_1-q_2-k).
\eeq
For the diagram Fig 2 $d$ this means inserting into the vertex the same
emission operator which acts in the single inclusive cross-section
corresponding  to Fig. 1 $c$.

Now we consider the second possibility when one (the quickest) gluon is
emitted  from the vertex and the other from the lower Pomeron (Fig. 2 $e$).
This time we have to return to the diagrams for reggeized gluon propagation
shown in Figs. 3-5. In contrast to the single inclusive cross-section studied
in Section 2, now we must take into account that at least one of the
Pomerons attached to gluons 12 or 34 (that from which the second gluon
is emitted) has to be cut.

As before we start from transitions from 2 to 4 gluons. Obviously the
diffractive cut passing through the center does not give any contribution,
since none of the lower Pomerons is cut. The double cut, passing through
the two Pomerons, will give two contributions, corresponding to the
observation of the left real gluon in the vertex
\[
-2g^4N_c\Big(P(1',2')V_lG(1',2'|1,2)P(3,4)\]\beq+
P(1,2)P(3',4')V_lG(3',4'|3,4)\Big)\otimes W(1,23,4|1k,234-k)D(1k)
\eeq
or the right real gluon in the vertex
\[
-2g^4N_c\Big(P(1',2')V_lG(1',2'|1,2)P(3,4)
\]\beq+
P(1,2)P(3',4')V_lG(3',4'|3,4)\Big)\otimes W(1,23,4|123k,4-k)D(4-k).
\eeq
In the single cut contributions the second gluon can only be emitted from the
cut lower Pomeron, the left one:
\beq
4g^4N_cP(1',2')V_lG(1',2'|1,2)P(3,4)
\otimes W(1,23,4|1k,234-k)D(1k)
\eeq
or the right one
\beq
4g^4N_c
P(1,2)P(3',4')V_lG(3',4'|3.4)\otimes W(1,23,4|123k,4-k)D(4-k).
\eeq
Due to symmetry under the interchange of gluons $(12)\leftrightarrow (34)$
contributions from the double and single cuts cancel. So transitions from
2 to 4 gluons give no contribution to the inclusive cross-section
corresponding to the diagram Fig. 2 $e$.

The same is true for transitions from 4 to 4 gluons of the type shown
in Fig. 5. The real gluon inside the vertex corresponds to only
the diffractive cut when the lower Pomerons are both uncut.

We are left with transitions from 3 to 4 gluons, Eq. (14), the first
term of which illustrated in Fig. 3.  Let us start from this term.
Again the diffractive cut gives no contribution. From the double cut we
have a contribution corresponding to the observed left real gluon in the
vertex:
\[
2g^3\sqrt{2N_c}\Big(P(1',2')V_lG(1',2'|1,2)P(3,4)
\]\beq
+
P(1,2)P(3',4')V_lG(3',4'|3.4)\Big)\otimes W
(1,2,3|1k,23-k)D_3^{(134)}(1k,23-k,4).
\eeq
The single cut contribution comes only from the cut Pomeron 12:
\beq
-4g^3\sqrt{2N_c}P(1',2')V_lG(1',2'|1,2)P(3,4)
\otimes W(1,2,3|1k,23-k)D_3^{(134)}(1k,23-k,4).
\eeq
In the sum we get
\[
2g^3\sqrt{2N_c}\Big(
P(1,2)P(3',4')V_lG(3',4'|3,4)
\]\beq
-P(1',2')V_lG(1',2'|1,2)P(3,4)\Big)\otimes W
(1,2,3|1k,23-k)D_3^{(134)}(1k,23-k,4).
\eeq
Due to assymetry between contributions from the same 12 or different 34
lower Pomerons they do not cancel and generate a non-zero inclusive
cross-section coresponding to Fig. 2 $e$. Below we list analogous
contributions from the rest of the terms in Eq. (14).
The second term gives
\[
2g^3\sqrt{2N_c}\Big(
P(3,4)P(1',2')V_lG(1',2'|1,2)
\]\beq
-P(3',4')V_lG(3',4'|3,4)P(1,2)\Big)
\otimes W(1,2,4|1k,24-k)D_3^{(134)}(1k,3,24-k).
\eeq
The third term  gives
\[
2g^3\sqrt{2N_c}\Big(
P(3,4)P(1',2')V_lG(1',2'|1,2)\]\beq
-P(3',4')V_lG(3',4'|3,4)P(1,2)\Big)
\otimes W(2,3,4|23k,4-k)D_3^{(124)}(1,23k,4-k)
\eeq
and the last one
\[
2g^3\sqrt{2N_c}\Big(
P(1,2)P(3',4')V_lG(3',4'|3,4)
\]\beq
-P(1',2')V_lG(1',2'|1,2)P(3,4)\Big)\otimes W
(1,3,4|13k,4-k)D_3^{(124)}(13k,2,4-k).
\eeq

Summing all the contributions and using the symmetries under
$1\leftrightarrow 2$
and $3\leftrightarrow 4$ we finally get the inclusive cross-section
\[
I^{(Z)}_1(k,l)=
2g^4N_cP(3,4)P(1',2')V_lG(1',2'|1,2)\]\beq
\Big\{W(1,2,3|1k,23-k)\Big(D(4)-D(23-k)\Big)-W(1,3,4|13k,4-k)
\Big(D(13k)-D(2)\Big)\Big\}+\Big(12\leftrightarrow 34\Big).
\eeq
Explicitly this cross-section has the form (passing to $\Phi$'s)
\[
I^{(Z)}_1(k,l)=
2g^2N_c\int\frac{d^2k_1}{(2\pi)^2}\frac{d^2k'_1}{(2\pi)^2}
\frac{d^2k_3}{(2\pi)^2}\Phi(3)\Phi(1'+l)V_lG(1'|1)
\Big\{\Big(\frac{k_1^2}{k^2(k_1+k)^2}
\]\[+\frac{k_3^2}{(k_1+k)^2(k_1+k-k_3)^2}
-\frac{(k_1-k_3)^2}{(k_1+k)^2(k_1+k-k_3)^2}\Big)\Big(
k_3^4P(k_3)-(k_1+k-k_3)^4P(k_1+k-k_3)\Big)\]\beq+\Big(1\leftrightarrow 3\Big)
\Big\}+\Big\{12\leftrightarrow 34\Big\}.
\eeq
Thus we get a non-trivial inclusive cross-section corresponding to
the diagram Fig. 2 $e$ which is naively prohibited by the AGK rules.
Note that the emission from the vertex operator is found to be different
from that in the diagram Fig. 2 $d$ allowed by the AGK rules.

\subsection{Transition to the vertex $V$}
As with the single inclusive cross-section we have to pass from the
vertex $Z$ to the vertex $V$ in the triple Pomeron diagrams Fig. 2 $a,b$
and also take into account the double Pomeron exchange diagram  Fig. 2 $c$.

We begin with the diagram Fig. 2 $a$ with both emissions from the upper
Pomeron. Transition to the vertex $V$ here is achieved in full similarity with
the single inclusive cross-section. The only difference in the derivation
is that the uppermost Pomeron has to be substituted according to (5).
As a result we get an additional contribution for the emission from
the vertex and upper pomeron (Fig 2 $d$), the new emission from the vertex
described by the Kovchegov-Tuchin operator:
\[
I^{(KT)}(k,l)=-\frac{g^2N_c}{\pi}\frac{1}{l^2}
\int\frac{d^2k_1}{(2\pi)^2}\frac{d^2k_3}{(2\pi)^2}
\frac{d^2k_5}{(2\pi)^2}
\]\beq
\Phi(1)\Phi(3)(k_1+k_3)^2
 (k_1+k_3+l)^2G(13l|5)V_kP(5+k).
\eeq

Now we pass to the diagram with two emissions from the lower Pomerons
Fig. 2 $b$. Here we use the fact that the gluon $k$ has its rapidity $y_1$
larger than the rapidity $y_2$ of the gluon $l$. Accordingly we use
the identity
\beq
G_{y-y_2}(3'',4''|3,4)=G_{y_1-y_2}(3'',4''|3',4')\otimes G_{y-y_1}(3',4'|3,4)
\eeq
and present the part below the vertex $Z$ at rapidity $y$ in the form
\[
P_{y_1}(1',2')V_kG_{y-y_1}(1',2'|1,2)P_{y_2}(3'',4'')V_l
G_{y_1-y_2}(3'',4''|3',4')\otimes G_{y-y_1}(3',4'|3,4)
\]\beq =
P_{y_1}(1',2')V_kP_{y_2}(3'',4'')V_lG_{y_1-y_2}(3'',4''|3',4')
\otimes G_{4,y-y_1}(1',2',3',4'|1,2,3,4),
\eeq
where $G_4$ is the Green function for the four gluons.
After integration with the vertex and upper part we present the diagram
Fig. 2 $b$ as
\[
P_{y_1}(1',2')V_kP_{y_2}(3'',4'')V_lG_{y_1-y_2}(3'',4''|3',4')\]\beq
\otimes G_{4,y-y_1}(1',2',3',4'|1,2,3,4)\otimes Z(1,2,3,4|\bar{1},\bar{4})
D(\bar{1}).
\eeq

Now we recall the basic relation between the verteces $Z$ and $V$:
\beq
D_4=G_4\otimes D_4^{(0)}+G_4\otimes Z\otimes D=D_4^{(R)}+
G_4\otimes V\otimes D,
\eeq
wherefrom we find
\beq
G_4\otimes Z\otimes D=D_4^{(R)}+G_4\otimes V\otimes D-G_4\otimes D_4^{(0)}.
\eeq
Putting this into (60) we find three terms for the double inclusive
cross-section.
The second is just the
desired structure with two emissions from the lower Pomerons and vertex $V$:
\beq
P_{y_1}(1',2')V_kG_{y-y_1}(1'2'|1,2)P_{y_2}(3',4')V_l
G_{y-y_2}(3',4'|3,4)\otimes V(1,2,3,4|\bar{1},\bar{4})\otimes D(\bar{1}).
\eeq
The third corresponds to the emission from the double Pomeron exchange
with the minus sign:
\beq
-P_{y_1}(1',2')V_kP_{y_2}(3'',4'')V_lG_{y_1-y_2}(3'',4''|3',4')
\otimes G_{4,Y-y}(1',2',3',4'|1,2,3,4)\otimes D_4^{(0)}(1,2,3,4)
\eeq
This contribution will cancel the same contribution from the diagram
Fig. 2 $c$.

Finally from the first term we shall have a new contribution to the
emission from the
vertex and a lower Pomeron (Fig. 2 $e$) with a structure
\beq
P_{y_1}(1,2)V_kP_{y_2}(3',4')V_lG_{y_1-y_2}(3',4'|3,4)
\otimes D^{(R)}_{Y-y_1}(1,2,3,4).
\eeq
This contribution is subdomimant at high energies, since
the term $D^{(R)}$ in (62) is obviously subdominant relative to the rest two.
So strictly speaking we have to drop it just as we have done with a similar
term for the single inclusive cross-section (cf. Eq. (33)).
However it grows faster than the contributions coming from the emissions
from the upper Pomeron or the upper Pomeron and the vertex (Fig. 2 $a,d$).
So we study this term in some more detail, so much the more that it
has certain peculiarities.

Using the explicit expression of $D^{(R)}$ (see ~\cite{BW, bra3}),
(65) generates the cross-section
\beq
I_2(k,l)=-\frac{1}{2}g^2
P_{y_1}(1,2)V_kP_{y_2}(3',4')V_lG_{y_1-y_2}(3',4'|3,4)
\otimes \Big(\sum_{i=1}^4D_{Y-y_1}(i)-\sum_{i=2}^4D_{Y-y_1}(1i)\Big).
\eeq
All terms depending on the sum 34 give zero, since the Green function
$G(3',4'|3,4)$ vanishes when the two reggeons are located at the same point
in the configuration space. Using the symmetry in  $1\leftrightarrow 2$
and $3\leftrightarrow 4$ we get
\beq
I_2(k,l)=-g^2
P_{y_1}(1,2)V_kP_{y_2}(3',4')V_lG_{y_1-y_2}(3',4'|3,4)
\otimes \Big(D_{Y-y_1}(3)-D_{Y-y_1}(13)\Big).
\eeq
The first term gives (passing to $\Phi$'s)
\beq
I_2^{(1)}(k,l)=-\frac{g^2N_c}{\pi k^2}\int\frac{d^2k_1}{(2\pi)^2}
\Phi_{y_1}(1)\frac{k_1^2}{(k_1-k)^2}\int \frac{d^2k_3'}{(2\pi)^2}
\frac{d^2k_3}{(2\pi)^2}\Phi_{y_2}(3')V_lG_{y_1-y_2}(3'|3)
k_3^4P_{Y-y_1}(3).
\eeq
The integral over $k_3$ and $k'_3$ is just 1/2 of the single inclusive
cross-section from a single Pomeron exchange $J(l)$.
The first factor corresponds to the lower Pomeron 'opened' at the top
and integrated over the topmost gluons. Such a contribution
does not look very natural (see Fig. 7 $a$). Moreover the integral
over $k_1$ obviously diverges at $k_1=k$. Of course this divergence
is in fact spurious and arises only because we have neglected
the momentum transfer in the target, taking it at $t=0$. Its appearence
is a signal that this approximation is too crude: in fact the first
factor behaves as $-(1/y_1)\log t$ as $t\to 0$. Integration  with the
nuclear wave function will convert this behaviour into a factor
proportional to $(1/y_1)\log A$.

The second term in (67) gives a non-factorizable integral
\beq
I_2^{(2)}(k,l)=\frac{g^2N_c}{\pi k^2}\int\frac{d^2k_1}{(2\pi)^2}
\frac{d^2k_3'}{(2\pi)^2}\frac{d^2k_3}{(2\pi)^2}
\Phi_{y_1}(1)\Phi_{y_2}(3')V_lG_{y_1-y_2}(3'|3)
\frac{k_1^2(k_1+k_3)^4}{(k_1-k)^2}P_{Y-y_1}(13).
\eeq
Its structure is shown in Fig 7 $b$.
It also diverges at $k_1=k$ and so leads to terms proportional to
$(1/y_1)\log A$.

\begin{figure}
\unitlength=1mm
\special{em:linewidth 0.4pt}
\linethickness{0.4pt}
\begin{picture}(122.00,126.33)
\put(40.33,119.33){\circle{14.00}}
\put(95.33,119.33){\circle{14.00}}
\put(35.00,114.67){\line(0,-1){44.67}}
\put(46.00,114.33){\line(0,-1){44.67}}
\put(13.00,106.67){\line(0,-1){37.00}}
\put(24.67,106.67){\line(0,-1){37.33}}
\put(90.67,113.67){\line(0,-1){13.33}}
\put(100.33,114.00){\line(0,-1){13.67}}
\put(91.00,100.33){\line(-3,-5){17.00}}
\put(100.33,100.00){\line(-3,-5){17.00}}
\put(90.67,99.67){\line(2,-3){18.67}}
\put(100.33,100.00){\line(2,-3){19.00}}
\put(19.00,110.67){\line(-4,-3){6.00}}
\put(19.00,110.33){\line(3,-2){6.00}}
\put(7.67,106.00){\line(1,0){5.33}}
\put(25.00,106.00){\line(1,0){6.00}}
\put(13.00,101.33){\line(1,0){11.67}}
\put(13.00,94.67){\line(1,0){11.67}}
\put(13.00,88.33){\line(1,0){11.67}}
\put(13.00,81.67){\line(1,0){11.67}}
\put(29.33,91.33){\line(1,0){5.67}}
\put(46.00,91.33){\line(1,0){6.00}}
\put(35.00,106.00){\line(1,0){11.00}}
\put(35.00,100.33){\line(1,0){11.00}}
\put(35.33,95.33){\line(1,0){10.67}}
\put(35.00,85.67){\line(1,0){11.00}}
\put(35.33,80.00){\line(1,0){10.67}}
\put(76.33,85.00){\line(1,0){5.67}}
\put(91.67,85.00){\line(1,0){4.67}}
\put(79.67,81.00){\line(1,0){9.33}}
\put(77.33,77.00){\line(1,0){9.33}}
\put(99.33,87.00){\line(1,0){9.67}}
\put(102.33,82.67){\line(1,0){9.67}}
\put(98.67,78.33){\line(1,0){6.67}}
\put(115.00,78.33){\line(1,0){5.33}}
\put(107.33,75.00){\line(1,0){9.67}}
\put(6.00,108.00){\makebox(0,0)[cc]{k}}
\put(53.00,93.67){\makebox(0,0)[cc]{l}}
\put(75.00,87.00){\makebox(0,0)[cc]{k}}
\put(122.00,80.00){\makebox(0,0)[cc]{l}}
\put(30.00,57.33){\makebox(0,0)[cc]{a}}
\put(96.67,57.33){\makebox(0,0)[cc]{b}}
\end{picture}
\vspace{-5.5 cm}
\caption{Illustration of  terms $I_2^{(1)}$ and $I_2^{(2)}$,
Eqs. (68), (69)}
\label{fig7}
\end{figure}
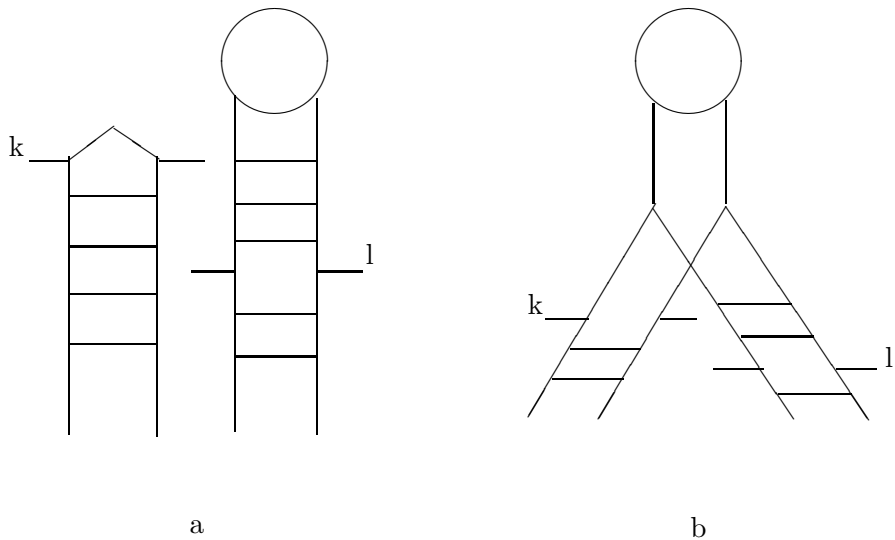

\section{Conclusions}
We have studied the single and double inclusive cross-sections by inspection
of the reggeized gluon diagrams in the Lipatov-Bartels formalism, as
an alternative to the colour dipole approach employed in
~\cite{KT, JK}. As a result we have found the cross-sections to be much
more complicated than obtained both in the latter approach and by
the naive application of the AGK rules in ~\cite{bra2}.
Various terms corresponding to the emission of gluon from the
triple Pomeron vertex have been discovered. Among them the term
derived in ~\cite{KT}, which in their approach comes by the change
from the quark dipole to the gluon dipole, in our treatment emerges
as a result of transition from the original diffractive vertex $Z$ to the
effective vertex $V$ used in the BK equation. However it does not exhaust
all the contributions to the emission from the vertex, which are numerous
and rather complicated in structure.

In the double inclusive cross-section we have found terms which correspond
to the emission of one gluon from the vertex and the other from one
of the two Pomerons immediately below the vertex. Such terms are not
expected by the AGK rules provided the cut vertex does not depend on
the way it is cut, which is usually assumed to be valid. Our results
show that the latter is not so. So while strictly speaking one cannot
state that the AGK rules are violated, their naive application does not
seem to hold. Still it is to be noted that we have not found contributions
corresponding to emission from Pomerons both above and below the vertex,
as in ~\cite{JK}, which would indeed strongly violate the AGK rules.
Also all found terms in the double inclusive cross-section with
somewhat unexpected structure  are subdominant at high energies and
their inclusion seems to be questionable.

We have to acknowledge that our derivation has been rather euristic.
It has been based on the study of reggeized gluon diagrams
for the triple discontinuity of the 4-gluon amplitude and assumes that the
AGK rules are fulfilled for different discontinuities.
Analysis of the gluon content of the triple gluon vertex has been performed
using its diagrammatic representation, by the visual inspection of the way
the real gluons are cut in the vertex. Both of these ingredients
of our approach have not be rigorously proven,
that is interpreted in terms of relevant production amplitudes.
So one has to take our derivation
with a certain dose of caution. The complicated character of the
contributions to emission from the vertex does not correspond to our
expectations for something so fundamental. However it does not seem
that one can avoid certain
extra terms, apart from the easily located KT term, for emission
from the vertex, since already in the high-mass diffraction
such terms obviously appear (see ~\cite{BW}).

\section{Acknowledgements}
The author is deeply indebted to J.Bartels, Yu.Kovchegov and G.P.Vacca for
numerous helpful discussions and critical comments. This work has been
supported by the NATO grant PST.CLG. 980287.

\section{Appendix 1. Inclusive cross-section from a single Pomeron exchange}
In this Appendix we derive the inclusive cross-section for a single
Pomeron exchange with emphasis on the numerical coefficient, important
for the comparison with other contributions. We follow normalizations
of Lipatov in ~\cite{lip}. Our basic equations will be the unitarity
relation for the reggeized gluon ladder together with the form
of the multiregge amplitude for production of $n$ real gluons.
As compared to ~\cite{lip} we make the following changes. We lift the
integration over the last reggeized gluon attached to the
target (momentum $q_{n+1}$) and consider the amplitude at fixed $q_{n+1}$.
Then the number of momentum integrations $n$ will match the number of
produced real gluons in the integmediate states. This allows to include
each factor $1/(4\pi)$ in the phase volume into the gluon interaction
substituting in it $g^2\to g^2/(2\pi)$ and absorbing 1/2 into the sum
over polarizations
\beq
(1/2)\sum_{\mu}C_\mu(q_1,q_2)C^{\mu}(q_1,q_2)=U(q_1,q_2).
\eeq
With these assumptions, the forward unitarity relation becomes
\beq
{\rm Im}\, A_n=\frac{\pi}{s}g^2\left(\frac{g^2N_c}{2\pi}\right)^{n}
\int\prod_{i=1}^{n}\frac{d^2q_i}{(2\pi)^2}
\int\prod_{i=1}^{n+1}ds_{i,i-1}
\delta(\prod_{i=1}^{n+1}s_{i,i-1}-s\prod_{i=1}^{n}p_i^2)
|B_n(q_i)|^2,
\eeq
where the intermediate real gluon have momenta $p_i=q_i-q_{i+1}$, $i=1,n$ and the
production amplitude $B_n$ has a multiregge form
\beq
B_n=2s \frac{s_{n+1,n}^{\omega_{n+1}}}{t_{n+1}}
\prod_{i=1}^{n}\frac{s_{i,i-1}^{\omega_i}}{t_i}\Big(e_iC(q_{i+1},q_i)\Big).
\eeq
Here $t_i=-q_i^2$, $e_i$ are the gluon polarization vectors,
$\omega_i=\omega(t_i)$ and $\omega(t)$ is the gluon Regge
trajectory which can be found in ~\cite{lip} together with the explicit
expression for vectors $C$. In (71) and (72) we suppressed all the colour
indeces,  summation over which is done in the standard way.

Our aim is to factorize the unitarity relation into the product of two
independent ones for parts of the ladder above and below some particular
real gluon with momentum $p_m$. To this end we first factorize the
production amplitudes introducing
\beq
B^{(1)}_{m-1}=2s_1 \frac{s_{m,m-1}^{\omega_{m}}}{t_m}
\prod_{i=1}^{m-1}\frac{s_{i,i-1}^{\omega_i}}{t_i}\Big(e_iC(q_{i+1},q_i)\Big)
\eeq
and
\beq
B^{(2)}_{n-m}=2s_1 \frac{s_{n+1,n}^{\omega_{n+1}}}{t_{n+1}}
\prod_{i=m+1}^{n}\frac{s_{i,i-1}^{\omega_i}}{t_i}\Big(e_iC(q_{i+1},q_i)\Big),
\eeq
with $s_1s_2=sp_m^2$. Obviously
\beq
B_n=\frac{s}{2s_1s_2}B^{(1)}_{m-1}B^{(2)}_{n-m}(e_mC(q_{m+1},q_m))
\eeq

Integration over the intermediate momenta of the square moduli
 of the amplitudes $B_{m-1}^{(1)}$ and
$B^{(2)}$ generates unitarity relations for the corresponding
elastic amplitudes:
\beq
{\rm Im}\, A^{(1)}_{m-1}=\frac{\pi}{s_1}g^2
\left(\frac{g^2N_c}{2\pi}\right)^{m-1}
\int\prod_{i=1}^{m-1}\frac{d^2q_i}{(2\pi)^2}
\int\prod_{i=1}^{m}ds_{i,i-1}
\delta(\prod_{i=1}^{m}s_{i,i-1}-s_1\prod_{i=1}^{m-1}p_i^2)
|B^{(1)}_{m-1}(q_i)|^2
\eeq
and
\[
{\rm Im}\, A^{(2)}_{n-m}=\frac{\pi}{s}\left(\frac{g^2N_c}{2\pi}\right)^{n-m}
\int\prod_{i=m+2}^{n}\frac{d^2q_i}{(2\pi)^2}\]\beq
\int\prod_{i=m+1}^{n+1}ds_{i,i-1}
\delta(\prod_{i=m+1}^{n+1}s_{i,i-1}-s_2\prod_{i=m+1}^{n}p_i^2)
|B^{(2)}_{n-m}(q_i)|^2.
\eeq

Now we turn to the overall unitarity relation (71). We present in it
\beq
\delta(\prod_{i=1}^{n+1}s_{i,i-1}-s\prod_{i=1}^{n}p_i^2)=
\int ds_1ds_2\delta(s_1s_2-sp_m^2)
\delta(\prod_{i=1}^{m}s_{i,i-1}-s_1\prod_{i=1}^{m-1}p_i^2)
\delta(\prod_{i=m+1}^{n+1}s_{i,i-1}-s_2\prod_{i=m+1}^{n}p_i^2).
\eeq
Then using the unitarity relations (76) and (77) and summing over the
polarizations of the distinguished gluon $p_m$ we obtain
\beq
{\rm Im}\, A_n=\frac{g^2N_c}{8\pi^2}\int ds_1ds_2\frac{s}{s_1s_2}
\int\frac{d^2q_m}{(2\pi)^2}\frac{d^2q_{m+1}}{(2\pi)^2}
\delta (s_1s_2-sp_m^2){\rm Im}\,A^{(1)}_{m-1} {\rm Im}\, A^{(2)}_{n-m}
U(q_{m+1},q_m).
\eeq

In the Lipatov normalization
\beq
{\rm Im}\,A=2sP
\eeq
So relation (79) in terms of Pomerons aquires the form
\beq
P_n=\frac{g^2N_c}{4\pi^2}\int ds_1ds_2
\int\frac{d^2q_m}{(2\pi)^2}\frac{d^2q_{m+1}}{(2\pi)^2}
\delta (s_1s_2-sp_m^2)P_{m-1}(q_m) P_{n-m}(q_{m+1})
U(q_{m+1},q_m).
\eeq
The inclusive cross-section results by summing over all numbers of gluons
in the two Pomerons on the right-hand side of (81),
fixing momentum $p_m=k$ and doubling the contribution according to (80)
\beq
d\sigma=4\alpha_sN_c\int ds_1ds_2
\frac{d^2k}{(2\pi)^3}\int\frac{d^2q}{(2\pi)^2}
\delta (s_1s_2-sk^2)P(q) P(q-k)U(q-k,q).
\eeq

Thus the final recipe for the inclusive cross-section is to introduce
the operator
\beq
V_k=2\alpha_sN_cU(q',q)\delta^2(q-q'-k),
\eeq
multiply the result by 2
and the resulting cross-section is
\beq
I(k)=\frac{(2\pi)^3d\sigma}{dyd^2k}.
\eeq

\section{Appendix 2. Explicit expressions for $I^{(Z)}$}
To find the explicit expressions for the inclusive cross-section
$I^{(Z)}(k)$ given by Eq. (25) we shall use the following relations
between the Pomerons $P(k)$, Pomerons in the nucleus $\Phi(k)$,
function $\phi(k)$ which satisfies the non-linear BFKL equation and
gluon density defined as
$ h(k)=k^2\nabla_k^2\phi(k)$:
\beq
g^2P(k)\to\Phi(k)=-
2\pi\nabla_k^2\phi(k)=(2\pi)^2\delta^2(k)-2\pi\frac{h(k)}{k^2},
\eeq
\beq
P(k)=-2\pi \frac{h_0(k)}{k^2}.
\eeq
We shall also use $\Phi(r=0)=P(r=0)=0$ in the configuration space
and the relation
\beq
\Delta\ln r=2\pi\delta^2(r).
\eeq

Armed with these relations we study separate terms in (25).
The first one (with $D(3-k)$) has the form
\[
I^{(Z)}_1=g^2N_c\int\frac{d^2k_1}{(2\pi)^2}
\frac{d^2k_3}{(2\pi)^2}\Phi(k_1)\Phi(k_3)P(k_3-k)(k_3-k)^4\]\beq
\Big\{\frac{k_1^2}{k^2(k_1-k)^2}+\frac{k_3^2}{k^2(k_3-k)^2}-
\frac{(k_1-k_3)^2}{(k_1-k)^2(k_3-k)^2}\Big\}.
\eeq
The bracket vanishes if $k_1=0$ or $k_3=0$. So in principle we can
substitute $\Phi$'s
and $P$ by gluon densities neglecting the $\delta$ term in (85). However
for this particular term this is not convenient.
The second term in the brackets gives zero, since integration over $k_1$
then leads to $\Phi(r_1=0)$. The rest two terms can be rearranged to exhibit
absence of infrared divergence to obtain
\beq
I^{(Z)}_1=g^2N_c\int \frac{d^2k_1}{(2\pi)^2}
\frac{d^2k_3}{(2\pi)^2}\Phi(k_1)\Phi(k_3)P(k_3-k)(k_3-k)^2
\Big\{\frac{k_1^2-k^2}{k^2(k_1-k)^2}+
2\frac{(\bf{k}_1-\bf{k})(\bf{k}_3-\bf{k})}{(k_1-k)^2(k_3-k)^2}\Big\}.
\eeq
Each of the two terms obviously factorizes in two integrals over $k_1$
and $k_3$.

The first term in the bracket contains two integrals:
\beq
\int\frac{d^2k_1}{(2\pi)^2}\Phi(k_1)\frac{k_1^2-k^2}{k^2(k_1-k)^2}=
\frac{2}{k^2}{\bf k}\nabla_k\phi(k)=\frac{2}{k}\phi'(k)
\eeq
and
\beq
\int\frac{d^2k_3}{(2\pi)^2}\Phi(k_3)P(k_3-k)
(k_3-k)^4= \int d^2re^{ikr}\Phi(r)\nabla^4P(r)\equiv X_1(k).
\eeq

In the second term the integral over $k_1$  can
be presented as
\beq
\nabla_k\int \frac{d^2k_1}{2\pi}\nabla^2_1\phi(k_1)\ln(k_1-k)=
\nabla_k\phi(k),
\eeq
where we have integrated by parts and used the relation (87).
The integral over $k_3$ can be written as
\beq
-\frac{1}{4}\int\frac{d^2k_3}{(2\pi)^2}\Phi(k_3)P(k_3-k)
\nabla_k(k_3-k)^4=
\frac{1}{4}\int d^2r {\bf r}\sin {\bf kr}\Phi(r)\nabla^4P(r)
\equiv {\bf k}X_2(k).
\eeq
Our final expession for this part is
\beq
I^{(Z)}_1(k)
=2g^2N_c\phi'(k)\Big(\frac{1}{k}X_1(k)+kX_2(k)\Big).
\eeq

The second term (with $D(4)$) differs from the first by the sign and change
$(k_3-k)^4P(k_3-k)\to k_3^4P(k_3)$. Obviously the factorization is
preserved with the integrals over $k_3$ changed. Now
\beq
\int\frac{d^2k_3}{(2\pi)^2}\Phi(k_3)P(k_3)
k_3^4=\int d^2k_3h(k_3)h_0(k_3)=\int d^2r\Phi(r)\nabla^4P(r)
\equiv X_3
\eeq
(it does not depend on $k$)
and the second integral
\beq
-\nabla_k\int\frac{d^2k_3}{(2\pi)^2}k_3^4\Phi(k_3)P(k_3)\ln(k_3-k)
=-\nabla_k\int d^2k_3h(k_3)h_0(k_3)\ln(k_3-k)\equiv{\bf k}X_4(k).
\eeq
So we find
\beq
I^{(Z)}_2(k)
=-2g^2N_c\phi'(k)\Big(\frac{1}{k}X_3+kX_4(k)\Big).
\eeq

The third and fourth terms (with $D(2k)$ and $D(1)$) differ from the
first two ones by the interchange of the two lower Pomerons and the
sign of $k$. Since the cross-section is obviously independent of this sign
and symmetric in the lower pomerons, the third and fourth terms give the
same contribution as the first two, so that (94) and (97) have to be doubled.

The fifth term (with $D(12k)$ is
\[
I^{(Z)}_5(k)=g^2N_ck^4P(k)
\int\frac{d^2k_1}{(2\pi)^2}\frac{d^2k_3}{(2\pi)^2}\Phi(1)\Phi(3)
\]\beq\Big\{\frac{k_1^2-k^2}{k^2(k_1-k)^2}+\frac{k_3^2-k^2}{k^2(k_3-k)^2}+
2\frac{(\bf{k}_1-\bf{k})(\bf{k}_3-\bf{k})}{(k_1-k)^2(k_3-k)^2}\Big\},
\eeq
where we transformed the initial $W(1,23,4|12k,34-5)$ similarly
to the transition from (88) to (89). The first two terms in the bracket do
not give any contribution, so that we are left with the last one,
which obviously factorizes in two identical integrals (92). So we find
\beq
I^{(Z)}_5(k)=2g^2N_ck^4P(k)\Big(\phi'(k)\Big)^2.
\eeq

Finally we have contributions from the sixth and seventh terms.
They do not factorize and we just rewrite their sum in terms of $h$ and $h_0$:
\[
I^{(Z)}_{6+7}(k)=-\frac{g^2N_c}{2\pi}
\int\frac{d^2k_1}{k_1^2}\frac{d^2k_3}{k_3^2}h(k_1)h(k_3)
\Big((k_1+k_3-k)^2h_0(k_1+k_3-k)-(k_1-k_3)^2h_0(k_1-k_3) \Big)
\]\beq
\Big\{\frac{k_1^2-k^2}{k^2(k_1-k)^2}+\frac{k_3^2-k^2}{k^2(k_3-k)^2}+
2\frac{(\bf{k}_1-\bf{k})(\bf{k}_3-\bf{k})}{(k_1-k)^2(k_3-k)^2}\Big\}.
\eeq

The final inclusive cross-section $I^{(Z)}$ from the vertex $Z$ is
\beq
I^{(Z)}=2I^{(Z)}_1+2I^{(Z)}_2 +I^{(Z)}_5+I^{(Z)}_{6+7}.
\eeq



\begin{thebibliography}{99}
%
\bibitem{kovch} Yu.V.Kovchegov, Phys. Rev {\bf D 60} (1999) 034008;
{\bf D 61} (2000) 074018.
%
\bibitem {bra1} M.A.Braun, Eur. Phys. J. {\bf C 16} (2000) 337.
%
\bibitem {bra2} M.A.Braun, Phys. Lett. {\bf B 483} (2000) 115.
%

\bibitem{KT} Yu.V.Kovchegov and K.Tuchin, Phys. Rev. {\bf D 65} (2002) 074026.
%
\bibitem{JK} J.Jalilian-Marian and Yu.V.Kovchegov, Phys. Rev. {\bf D 70}
(2004) 114012.
%
\bibitem{MP} A.H.Mueller and B.Patel, Nucl. Phys. {\bf B 425} (1994) 471.
%
\bibitem{BW} J.Bartels and M.Wuesthoff, Z.Phys. {\bf C 66} (1995) 157.
%
\bibitem{BV} M.A.Braun and G.P.Vacca, Eur. Phys. J. {\bf C 6} (1999) 147.
%
\bibitem{BT} M.A.Braun and D.Treleani, Eur. Phys. J. {\bf C 4} (1998) 685.
%
\bibitem {bra3} M.A.Braun, Eur. Phys. J. {\bf C 6} (1999) 321.
%
\bibitem{lip} L.N.Lipatov, in "Perturbative QCD", ed. A.H.Mueller
   (world Sci., Singapore, 1989).
%
\end{thebibliography}
\end{document}